\documentstyle[emulateapj, epsf]{article}

\lefthead{Larkin et al.}
\righthead{Distant Galaxies with Keck AO}
\begin{document}

\submitted{Accepted to PASP, August 23, 2000}

\title{Exploring the Structure of Distant Galaxies with Adaptive Optics on
the Keck-II Telescope}

\author{J. E. LARKIN$^1$, T. M. GLASSMAN$^1$, P. WIZINOWICH$^2$,
D. S. ACTON$^2$, O. LAI$^{2,3}$, A. V. FILIPPENKO$^4$, A. L. COIL$^4$,
AND T. MATHESON$^4$}

\affil{$^1$Dept. of Physics and Astronomy, University of California,
P.O. Box 951562, Los Angeles, CA 90095-1562,}
\affil{$^2$W.M. Keck Observatory, 65-1120 Mamalahoa Hwy., Kamuela, HI 96743,}
\affil{$^3$Canada-France-Hawaii Telescope Corp., P.O. Box 1597, Kamuela, HI
96743,}
\affil{$^4$Dept. of Astronomy, University of California, Berkeley, CA
94720-3411}

\authoremail{larkin@astro.ucla.edu, glassman@astro.ucla.edu,
peterw@keck.hawaii.edu, sacton@keck.hawaii.edu
alex@astron.berkeley.edu, acoil@astron.berkeley.edu,
tmatheson@astron.berkeley.edu}

\begin{abstract}

We report on the first observation of cosmologically distant field
galaxies with an high order Adaptive Optics (AO) system on an 8-10
meter class telescope. Two galaxies were observed at 1.6 $\mu$m at an
angular resolution as high as 50 milliarcsec using the AO system on
the Keck-II telescope. Radial profiles of both objects are consistent
with those of local spiral galaxies and are decomposed into a classic
exponential disk and a central bulge. A star-forming cluster or
companion galaxy as well as a compact core are detected in one of the
galaxies at a redshift of 0.37$\pm$0.05. We discuss possible
explanations for the core including a small bulge, a nuclear starburst,
or an active nucleus. The same galaxy shows a peak disk surface
brightness that is brighter than local disks of comparable size. These
observations demonstrate the power of AO to reveal details of the
morphology of distant faint galaxies and to explore galaxy evolution.

\end{abstract}


\section{Introduction}

It has only been very recently, with new facilities such as the Keck
telescopes and the \it Hubble Space Telescope (HST)\rm, that early
galaxy formation and galaxy evolution could be studied in detail.  The
high spatial resolution of the \it HST \rm has revealed that young
galaxies appear to be more numerous and smaller than nearby galaxies
(e.g., Phillips et al. 1997), and often have a distorted morphology
(e.g., Driver et al. 1998). Studies of Lyman break galaxies (e.g.,
Steidel 1999) and the Hubble Deep Fields have also revealed that star
formation was more intense at higher redshifts, although the star
formation rate as a function of time is still heavily debated. Taken
together, these attributes suggest that galaxies have gone through a
period of significant evolution since their formation. It has been
difficult, however, to obtain detailed answers about the growth of
galaxies and the rates of star formation and mergers. Also, still
unknown, is the prevalence of modest AGN activity at high redshifts.

Part of the difficulty in trying to answer these problems is the
extremely small angular sizes of distant galaxies. The exponential
scale lengths of spiral galaxies are predominantly in the range from 1
to 10 kpc (e.g., de Jong 1996).  Thus, by a redshift of $z$=0.5 all disks
have an angular scale length under an arcsecond. Bulges are typically
of order 10 times smaller than the disks. Consequently, very few
measurements have been made of the central spheroidal component of
galaxies at high redshifts. Several recent surveys have, however, been
able to measure disk properties at optical wavelengths and look for
evidence of disk evolution. In particular, Schade et al. (1996) as part
of the CNOC2 redshift survey, showed that the peak surface brightness
of disks increased 1.5 mag by $z\sim$0.8. Vogt (1999) argued that
within their own data (the DEEP Project) a similar effect could be
explained primarily by selection bias, and they concluded that disk
evolution is relatively mild.  Recent cold dark matter models by Mao,
Mo, \& White (1998) and Dalcanton, Spergel, \& Summers (1997) have
been used to model the rate of galaxy growth and have made predictions
that disk radii should decrease as (1+$z$)$^{-1}$ with roughly
constant $B$-band luminosity so that the intrinsic surface brightness
must increase. In all of these cases, the use of $B$-band magnitudes
has been a significant complication since the K-corrections and
mass-to-luminosity ratios at $B$ are poorly constrained. It would be
preferable to use infrared images, especially in the $H$ and $K$
bands.  These wavelengths trace the older stellar population even at
fairly high redshifts and are less affected by dust or recent star
formation. Unfortunately, \it HST \rm cannot currently observe at
these wavelengths and NICMOS images had insufficient angular
resolution to resolve most distant bulges.

Adaptive optics (AO) systems provide real-time compensation of the
blurring effects of the Earth's atmosphere (e.g., Beckers 1993).  Most
astronomical AO systems can be divided into two categories reflecting
their method of wavefront sensing and correction. The Keck AO system
falls in the Shack-Hartmann based category and uses a lenslet array to
measure the wavefront error at a location close to the conjugate of
the primary mirror.  A deformable mirror is also placed close to a
pupil plane and is used to correct for the measured aberrations. Since
the characteristic scale length of atmosphere turbulence (r$_0$
$\approx$ 10-40 cm) is much smaller than telescope diameters, a large
number of actuators is needed to produce good correction.  For the
Keck AO system (Wizinowich et al. 2000), 349 actuators are updated at
up to 660 Hz. Many of the most successful AO systems such as PUEO
(Rigaut et al. 1998) on the Canada-France-Hawaii Telescope use a
curvature sensor to measure wavefront error and have traditionally
used a bimorph mirror with a relatively small number of degrees of
freedom ($<$50). For several reasons including the use of avalanch
photodiodes and the need to divide the light into fewer elements,
curvature based systems can operate on fainter targets and have
historically been more successful on extragalactic targets.

In any AO system, only a fraction of the light is restored to the
diffraction-limited core, while the rest remains in an extended halo
roughly the size of the seeing disk. A parameter used to describe the
effectiveness of AO is the Strehl ratio. It is the ratio of the peak
intensity of the point-spread function (PSF) compared to the
theoretical peak intensity of a diffraction-limited source. A
higher-order system with many actuators (or degrees of freedom) will
normally result in a higher Strehl ratio than a lower-order system on
the same telescope. Typical Strehl ratios at 1.6 $\mu$m for the Keck
system are around 0.3 for the guide star and fall off with angular
separation. The Strehl ratio is typically below 0.1 at a distance of
$30''$ from the guide star, although there is significant variation with
atmospheric conditions.

The Keck system requires a guide star brighter than 12th mag at $R$
for full correction. Partial correction is possible on guide stars as
faint as 14th mag. This limitation of using bright sources for
wavefront references has severely limited the use of AO systems on
faint extragalactic targets. Our study of faint field galaxies is the
first of its kind and is only possible because of the high density of
galaxies on the sky. Deep infrared imaging surveys (e.g., Djorgovski
et al. 1995) have shown that in the $K$ band (2.2 $\mu$m), on average,
there are 2 galaxies brighter than $K$ = 20 mag within $20''$ of any
star.  These objects form a statistically unbiased set of galaxies at
high redshift and provide an excellent sample for studying galaxy
evolution.  At a typical redshift between 0.25 and 1.0 (corresponding
to look back times of (3-8)$\times$10$^9$ years ago) these objects are
seen at a critical time in galaxy development\footnote[5]{In this
paper we will use a cosmology with H$_0$=65 km/sec/Mpc,
$\Omega_M$=0.3, and $\Omega_\Lambda$=0.7 (Garnavich et
al. 1998). Assuming other models including an open universe with
q$_0$=0.1 does not significantly affect the results.}.  This
corresponds to just after the time that the Milky Way's disk appears
to have stabilized and when the oldest disk stars formed (Winget et
al. 1987). Recent evidence also indicates that star formation
throughout the Universe may have begun its decline to the present
levels at around $z=1$ (Madau, Pozzetti, \& Dickinson 1998).

\section{Observations}

Between 1998 September and 1999 July, 25 stars were imaged with a non-AO
infrared camera (NIRC; Matthews \& Soifer 1994) on the Keck-I
telescope. These images were used to provide a sample of galaxies
within $30''$ of potential guide stars. The stars were all
A-type stars with a visual magnitude between 8 and 12. Over 200
galaxies were selected in these fields down to a $K$-band magnitude of
21.5. Many of the fields also have faint off-axis stars that can be
used to measure the PSF in order to determine the level of correction
achieved. This is critical since the performance of AO systems can
vary significantly with position in the sky and brightness of the
guide star. These pre-AO images are detailed in Larkin \& Glassman
(1999).

The camera used for the AO observations was KCAM, which was developed
at UCLA by James Larkin and Ian McLean in coordination with Peter
Wizinowich and the AO development team.  It uses a NICMOS infrared
array detector with 256$\times$256 pixel elements. It is sensitive from 1 to
2.5 $\mu$m, although the current filter set limit it to $J$ (1.27
$\mu$m), $H$ (1.65 $\mu$m) and $K^\prime$ (2.15 $\mu$m). The plate
scale in the $H$-band is 0$\farcs$0176 $\pm$ 0$\farcs$0004 per pixel
yielding a total field of view of 4$\farcs$51 (Macintosh 2000).  This
plate scale was selected to Nyquist sample the $H$-band diffraction
limit.

On 1999 April 3, we observed the first faint field galaxy ever
examined with a Shack-Hartmann AO system. The galaxy is 18$''$
from a $V$=10.2 mag natural guide star (NGS) named PPM
127095 (R.A.=10$^h$05$^m$26.86$^s$,
dec=+16$^\circ$21$^\prime$50\farcs18, J2000). The galaxy (PPM
127095-8+16)\footnote[6]{We name each galaxy by the
position and proper motion (PPM) designation of the guide star
augmented by the offsets in arcseconds to the galaxy.} is 19.0 mag in
the $H$-band. Its infrared colors are $J-H$=0.9 mag and $H-K$=0.8
mag. Non-AO imaging marginally resolved the galaxy in good seeing
conditions (0\farcs5), but no morphological information was
available. A second star 17$\farcs$9 from the same NGS was used to
measure the PSF. Six exposures of five minutes each were taken on the
galaxy, with the PSF star observed before and after the galaxy
sequence. Between each of the six exposures, the galaxy was moved $2''$
in a square dither pattern. The total integration time was limited by
the demands of the AO commissioning program.

A second galaxy was observed in a similar manner in September
1999. The galaxy (PPM 114182+6+27) is $27''$ away from the 8th mag
A-type star PPM 114182 (R.A.=22$^h$13$^m$40.2$^s$,
dec=+21$^\circ$02$^\prime$58\farcs7, J2000) and has a total $H$-band
magnitude of 17.4 and an extent of roughly $3''$. Its infrared colors
are $J-H$=0.7 mag and $H-K$=0.1 mag.  In this particular case, the
Strehl ratio degraded rapidly during the observations from $\sim$0.2
to less than $\sim$0.1. In the final image, only the first three
frames (each 5 minutes) out of a total of six were utilized. The
variability of the PSF is discussed in the results section below. This
galaxy was also observed with the optical spectrograph LRIS (Oke et
al. 1995) on the Keck-II telescope on 1999 December 12. Two 10-minute
exposures were taken, but the first suffered from a large amount of
scattered light from the PPM star and was not used.

The images were reduced using standard techniques for non-AO images.
Sky frames were created by masking off the quadrant with the galaxy
from each frame and then they were medianed together without
positional offset.  For PPM~114182+6+27, frames with poor Strehl
ratios were used for creating median sky frames even if they weren't
used for the final image.  Flat field images were created from
observations of the twilight sky.

\section{Results}

Figure 1 shows the image of PPM~127095-8+16. The contours of the
galaxy are consistent with a disk galaxy with an inclination of
65$^\circ$. We removed the first-order effects of inclination by
rotating the image, making its major axis horizontal, then resampling
it in the vertical direction to produce roughly circular
isophotes. The radial profile was subsequently produced by azimuthally
averaging the image. This yields a high signal-to-noise ratio (S/N)
profile of the galaxy out to a surface brightness of $H\approx$22
mag/arcsec$^2$. Figure 1 also shows this radial profile for both the
galaxy and the PSF star.

{\plotfiddle{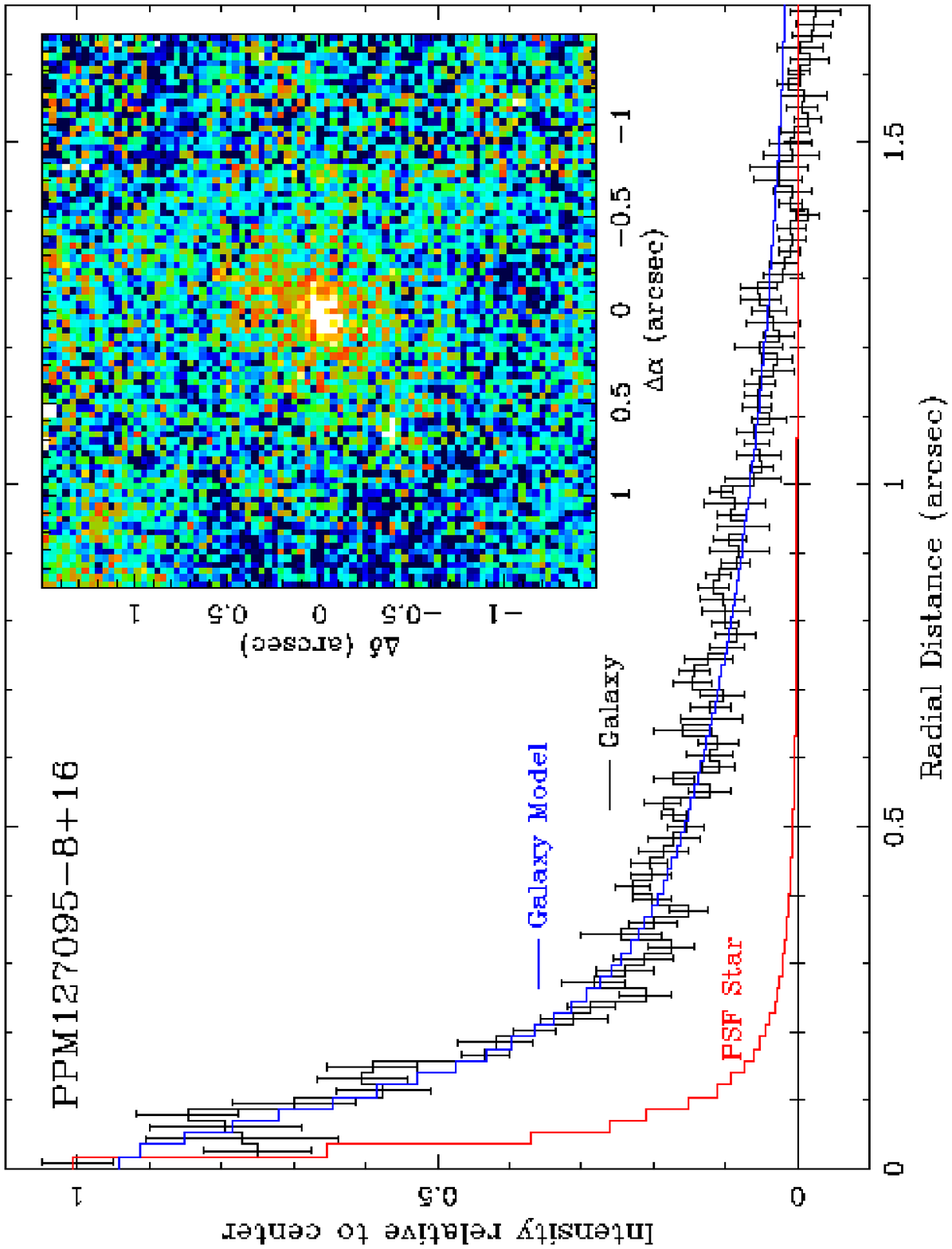}{3.2in}{270}{40}{40}{-28}{240}}
{\footnotesize Figure 1. - Faintest galaxy ever imaged with a high
order adaptive optics system ($H$=19.0 mag). This image of the galaxy
PPM~127095-8+16 has a resolution of 50 milliarcseconds and also
represents the highest resolution infrared image ever taken of a faint
field galaxy.}
\medskip

To decompose the galaxy into bulge and disk components, we created
2-dimensional model galaxies and convolved these models with the PSF
star image.  We used an exponential profile for both the bulge and the
disk and found that a de Vaucouleurs (1948) model for the bulge
produced significantly worse fits. The free parameters of the model
were the scale sizes and peak brightnesses of the exponential disk and
bulge plus an overall background offset. This offset is necessary
because the small field of view does not allow for an unambiguous
background measurement. Since we had good infrared photometry on the
galaxy in non-AO images, however, we were able to constrain the flux
in the galaxy within $3''$, which restricts the amount of background
fluctuation that is allowed in the model. Errors in the profile were
estimated empirically by performing the azimuthal averaging in four
angular bins, then determining the variation within the bins as a
function of radius. A grid search was used to determine these
parameters to the nearest two pixels for the disk and the nearest 0.2
pixel for the bulge. The best fit bulge scale length is 0\farcs042
with a 1$\sigma$ range from 0\farcs025 to 0\farcs056.  The disk scale
is 0\farcs63 with 1$\sigma$ limits of 0\farcs42 and 0\farcs81.  Of the
four galaxy parameters, the disk scale size is the most uncertain due
to a partial degeneracy with the background level. The measured peak
surface brightness of the disk is 19.6 mag/arcsec$^2$ at $H$ which
corresponds to $K$=18.8 mag/arcsec$^2$ (using non-AO colors to convert
$H$ to $K$). The peak surface brightness of the bulge is $H$=17.2
mag/arcsec$^2$. The radial profile of the best fit model galaxy is
also shown in Figure 1 as a dashed curve.

Figure 2 shows the image and radial profile of PPM~114182+6+27.
Unfortunately conditions were not as good as on the PPM~127095-8+16
source and the FWHM of the PSF star was just under 0\farcs1. Isophotal
contours of the galaxy are approximately circular so no inclination
correction is applied. The outer region of the radial profile is well
fit with an exponential disk with a scale length of 0\farcs92 with a
1$\sigma$ range of 0\farcs73 to 1\farcs23. The central region is best
fit by an exponential bulge with radial scale length of 0\farcs1.
However, if the disk component is subtracted from the galaxy, the
residual is only slightly wider than the PSF. Given the variability of
the psf during the observations of this galaxy it is possible that the
central region is not resolved. Using the point source model, the core
has an $H$-band magnitude of 20.8, $\sim$4\% of the total $H$-band
luminosity of the galaxy. This core would be undetectable without AO.

{\plotfiddle{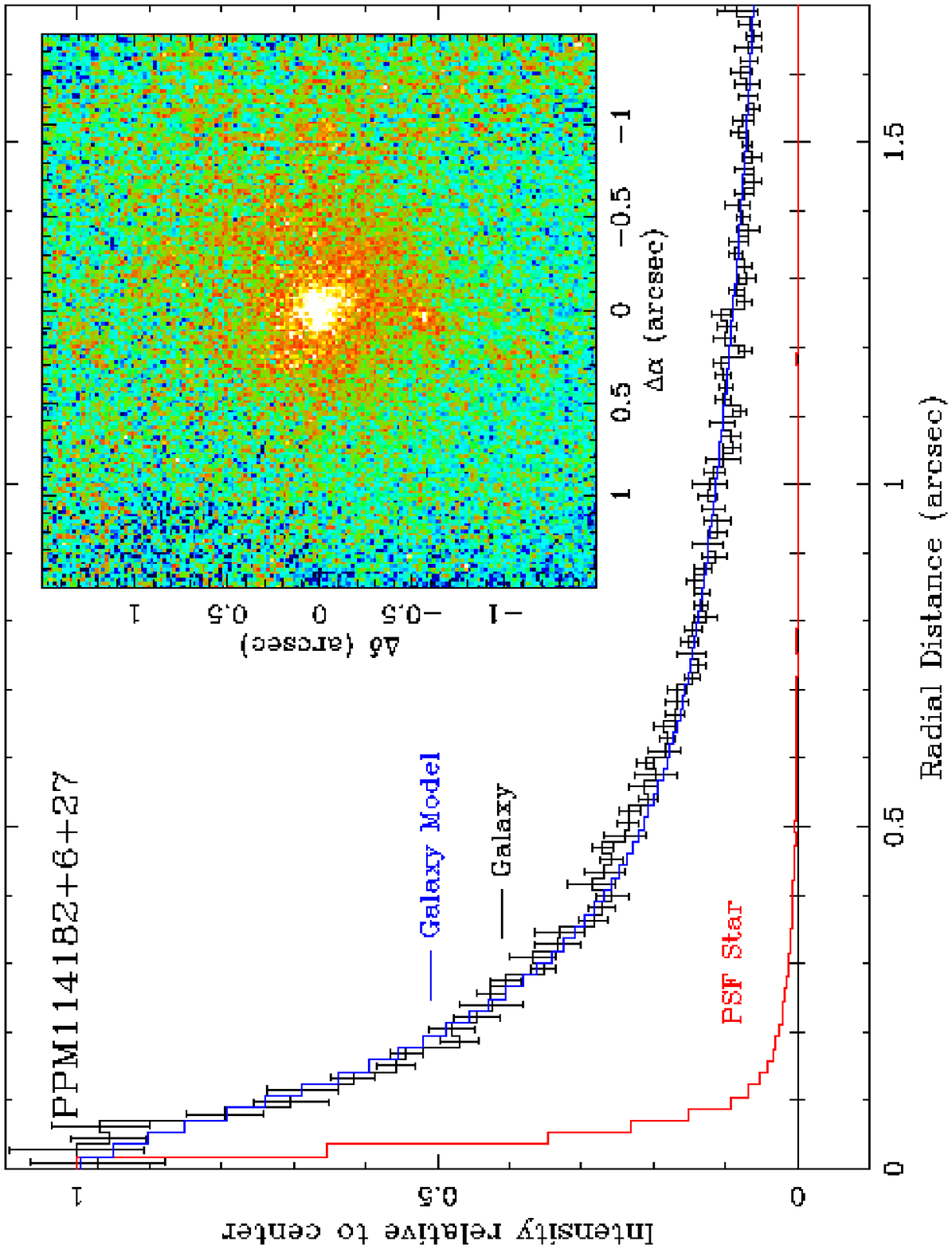}{3.2in}{270}{40}{40}{-28}{240}}
{\footnotesize Figure 2. - AO image of a galaxy at the epoch of the
formation of the Solar System. This galaxy (PPM~114182+6+27) shows
evidence for a disk but perhaps not a true bulge. Instead a central
unresolved core is found along with a point source 0$\farcs$6 south of
the center.}
\medskip

The LRIS spectrum of PPM~114182+6+27 (Figure 3) shows no spectral
lines but does have a significant break at 5500~\AA\, (observed
frame). If we interpret this as the 4000~\AA\, break we find good
agreement with a template spectrum at a redshift of 0.37$\pm$0.05. At
this redshift, the physical scale size of the disk is 5.1 kpc and the
scale size of the bulge is less than 300 pc (3 pixels). This
corresponds to a look back time of 4.4 Gyr, almost exactly the
estimated age of our own Solar System. Other possibilities for the
identification of the break would require a much higher redshift and
would be inconsistent with the observed surface brightness of the
disk. Regardless of redshift or cosmological model, the ratio of bulge
to disk scale size is below 0.12. For the galaxy PPM~127095-8+16,
no redshift is currently available.

\section{Discussion}

For direct comparison to the properties of local disk galaxies we
selected the de Jong (1996) sample of 86 nearly face-on spiral
galaxies. Its advantage is that it contains accurate bulge and disk
parameters in all major bands from $U$ to $K$. At a redshift of 0.37
the galaxy PPM~114182+6+27 has a luminosity distance of 2.1 Gpc giving
it an absolute $K$-band magnitude of -24.4. This is comparable to the
most luminous objects in the de Jong sample. Perhaps more importantly,
its peak disk surface brightness is 17.2 mag/arcsec$^2$ after
correcting for the (1+$z$)$^{-4}$ fading. This is again near the
maximum for all of the local sample of galaxies. Large disk galaxies
with comparable scale lengths (3-6 kpc) range from 17.5 to roughly
19.0 mag/arcsec$^2$ making the AO galaxy  brighter than
the brightest local galaxy of comparable size, and $\sim$1.0 mag
brighter than the average. The central source, however, has an
absolute magnitude of -20.8 in $H$ which is towards the middle of the
local distribution of bulges. Its size, however, is compact with a
scale length under 500 pc.

{\plotfiddle{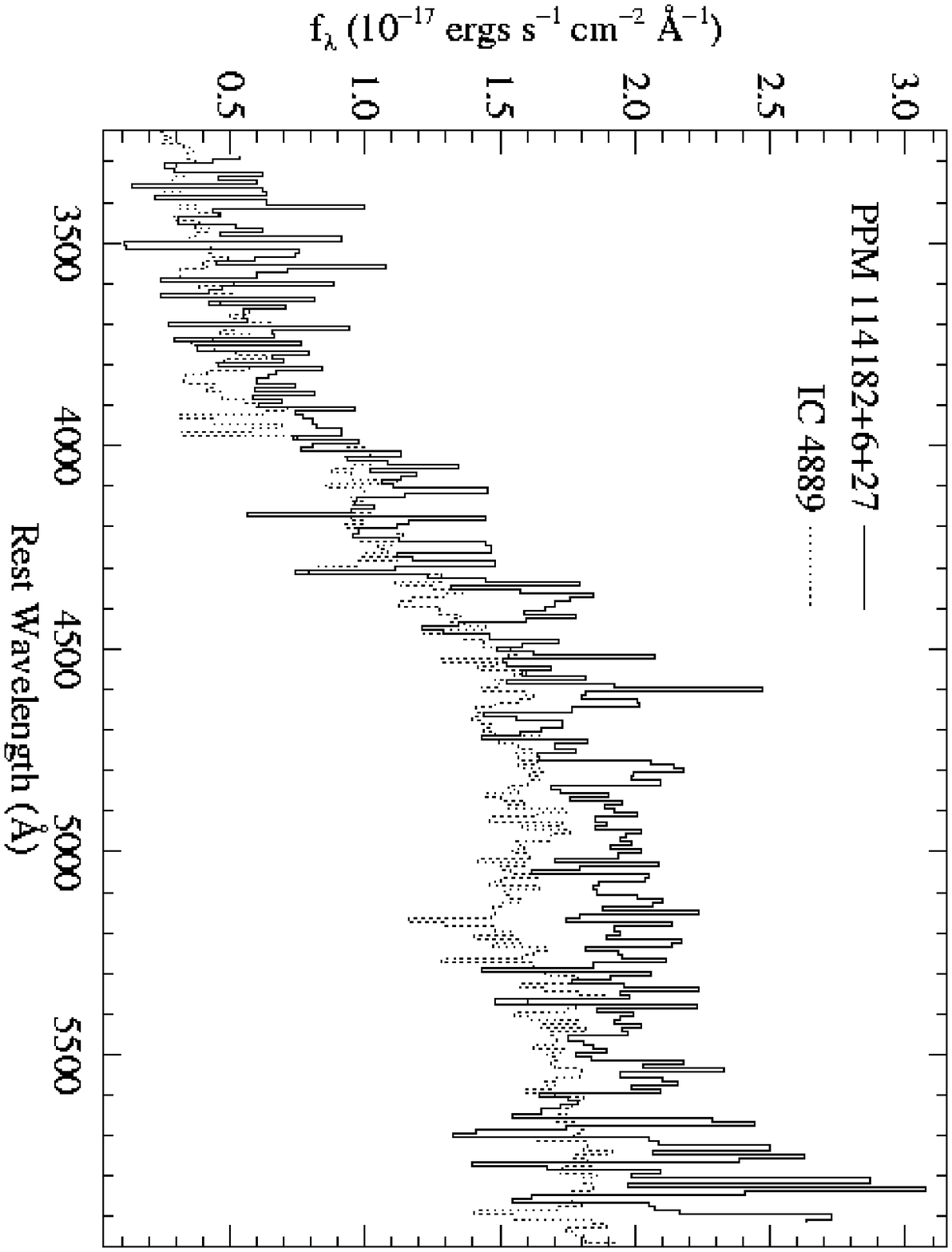}{3.2in}{90}{40}{40}{285}{-10}}
{\footnotesize Figure 3. - Optical spectrum of PPM~114182+6+27. It has
been shifted to match the spectrum of the local elliptical galaxy
IC~4889. The best shift corresponds to a redshift of 0.37$\pm$0.05
primarily based on the position of the spectral break at 4000~\AA. No
emission lines are visible.}
\medskip

One possible explanation for an unresolved core is a nuclear
starburst. Some local galaxies exhibit greatly enhanced star formation
within the central kpc, often due to tidal interactions. This
phenomenon might be expected to increase at high redshift as galaxy
separations decrease. We compared the bulge with the local starburst
galaxy NGC 1808. Tacconi-Garman et al. (1996) found that the central
750 pc of NGC 1808 has an absolute $K$-band magnitude of -22.3
corresponding to a star-formation rate of 11 M$_\odot$ per year. Using
$H-K$= 0.4 mag, this is only 0.1 mag different from our high-redshift
core.

The optical spectrum (Figure 3) does not contain a bright
[O\,II]~$\lambda$3727 emission line, however, and is generally
inconsistent with strong ongoing star formation.  If most of the star
formation in the disk occurred at an even earlier time, then the
[O\,II] line may be too faint for detection.  The central core, where
star formation may be active, is too weak to be detected in the
spectrum since it was taken without AO and thus includes most of the
galaxy.

Another possible contributor to the core would be an active
nucleus. Again, the core is probably too faint to have produced
detectable emission lines in the LRIS spectrum.  It is interesting to
consider that AO spectroscopy may be one of the most sensitive methods
for detecting faint AGNs at high redshifts since most of the starlight
is rejected.

The AO image of PPM 114182+6+27 (Figure 2) also shows a point source
0\farcs6 south of the nucleus, with an $H$-band magnitude of 22. The
knot then has an absolute $H$-band magnitude of -20 and may be a very
luminous star-forming region. Similar ``super star clusters'' are seen
in local galaxies, but this would be among the most extreme examples
and is about 2 mag brighter than the brightest knot in the interacting
galaxies NGC 6090 (Dinshaw et al. 1999). It is likely that the knot in
PPM~114182+6+27 is composed of many such regions and may even be a
second galactic nucleus indicating an ongoing merger or simply a
companion galaxy seen in projection against the disk. If it is a star
forming knot and others like this one prove to be common, then it may
indicate that star formation at high redshifts was often in giant
superclusters.

\section{Conclusions}

We have presented new near-diffraction-limited images of
cosmologically distant field galaxies taken with the Keck AO
system. Even these preliminary images show significant detail and
provide morphological information on scales as small as
0\farcs05. This gives us the first ability to accurately study
morphological evolution at cosmological distances from 8-10 meter
telescopes. Both of the objects presented here show a smooth
exponential disk and a bulge with a scale size at or below 0.1 of the
disk, consistent with local objects. One of the galaxies has a large
star-forming region or possibly a companion galaxy and an elevated
peak surface brightness.

We have recently obtained similar images of
over 10 additional galaxies and more observations are
scheduled. We soon hope to produce a statistically meaningful population
study of these objects. Also within the next year, with new
spectrographic capabilities and infrared colors, we plan to search for
the presence of active nuclei, study the stellar population ages, and
measure the rotation curves of these cosmologically young objects.

\acknowledgements
We would like to acknowledge the contributions of the members of the
Keck AO team including the wavefront controller team at LLNL.  We also
want to thank the Keck Observatory staff, especially Director Fred
Chaffee and our night assistants. The engineering science camera was
provided by UCLA and was originally constructed by Ian McLean.  Ian
McLean, Andrea Ghez, and Alycia Weinberger also provided useful
discussions and support on preliminary observing.  Funding for the
project came in part from the new NSF Center for Adaptive Optics under
the leadership of Jerry Nelson. A.V.F. is supported by NSF grants
AST-9417213 and AST-9987438, and A.L.C. acknowledges an NSF Graduate
Research Fellowship. Data presented herein were obtained at the
W.M. Keck Observatory, which is operated as a scientific partnership
among the California Institute of Technology, the University of
California, and the National Aeronautics and Space Administration.
The Observatory was made possible by the generous financial support of
the W.M. Keck Foundation.

\end{document}